\documentstyle[12pt]{article}

\newcommand{\hc}{\hbox {h.c.}}
\newcommand{\re}{\hbox {Re}}
\newcommand{\im}{\hbox {Im}}

\newcommand{\gev}{\hbox {GeV}}
\newcommand{\etal}{\hbox{et al.}}
\newcommand{\prdj}[1]{{ \it Phys.~Rev.}~{\bf D{#1}}}
\newcommand{\prlj}[1]{{ \it Phys.~Rev.~Lett.}~{\bf {#1}}}
\newcommand{\plbj}[1]{{ \it Phys.~Lett.}~{\bf {#1B}}}
\newcommand{\npbj}[1]{{ \it Nucl.~Phys.}~{\bf B{#1}}}
\newcommand{\ptpj}[1]{{ \it Prog.~Theor.~Phys.}~{\bf {#1}}}
\newcommand{\zfpj}[1]{{ \it Z.~Phys.}~{\bf C{#1}}}
\newcommand{\beq}{\begin{equation}}
\newcommand{\eeq}{\end{equation}}
\newcommand{\bea}{\begin{eqnarray}}
\newcommand{\eea}{\end{eqnarray}}
\newcommand{\baa}{\begin{array}}
\newcommand{\eaa}{\end{array}}
\newcommand{\ra} {\rightarrow}
\newcommand{\ttbar}{t\bar{t}}
\newcommand{\tanb} {\tan \beta}
\newcommand{\twddec} {t\rightarrow W^+ d_k}
\newcommand{\twbdec} {t\rightarrow W^+ b}
\newcommand{\tbwddec} {\bar{t} \rightarrow W^- \bar{d}_k}
\newcommand{\tbwbdec} {\bar{t} \rightarrow W^- \bar{b}}
\newcommand{\thddec}{t \ra H^+ d_k}

\newcommand{\tbar} {\bar{t}}

\newcommand{\wpp} {W^+}
\newcommand{\st} {\tilde{t}}
\newcommand{\glu} {\lambda}
\newcommand{\tsgdec} {t \ra \st \glu}
\newcommand{\mt}{m_t}
\newcommand{\mts}{m_t^2}
\newcommand{\mj}{m_j}
\newcommand{\mjs}{m_j^2}

\newcommand{\mis}{m_i^2}
\newcommand{\mw} {m_W}
\newcommand{\mws} {m_W^2}
\newcommand{\mh} {m_H}

\newcommand{\tbsgdec} {\tbar \ra \st^\dagger \bar{\glu}}
\newcommand{\imag}{\im(V^\ast_{tk}V^{}_{ti}V^\ast_{ji}V^{}_{jk})}
\begin{document}
\begin{flushright}
CERN-TH.6941/93\\
\hfill  hep-ph/9310286\\
\hfill UICHEP-TH/93-15
\end{flushright}
\vspace*{2cm}
\begin{center}
{\large{\bf The Decay Rate Asymmetry of the Top Quark}\\
\vspace*{2cm}
Bohdan Grz\c{a}dkowski\footnote{E-mail:
{\tt bohdang@cernvm.}\\
\noindent Address after 1 October 1993:
        Institute for Theoretical Physics,
        University of Warsaw,
        Ho\.{z}a 69, PL-00-681 Warsaw, Poland. }\\
\vspace*{.5cm}
Wai-Yee Keung\footnote{E-mail: {\tt keung@fnal.}\\
\noindent Address after 1 August 1993:
        Physics Department, M/C 273
        University of Illinois at Chicago,
        IL 60607--7059,
        USA. }\\
\vspace*{1cm}
       Theory Division, CERN, CH-1211 Geneva 23, Switzerland }\\
\vspace*{3cm}

{\bf Abstract} \\
\end{center}
The asymmetries
${\cal A}_k \equiv [\Gamma(\twddec) -\Gamma(\tbwddec)]
/[\Gamma(\twddec) + \Gamma(\tbwddec)]$
in the partial widths of the top quark decays are discussed
within the Standard Model (SM), the Two-Higgs-Doublet Model (2HDM) and
supersymmetric extensions of the SM (SSM). The leading contributions to
these asymmetries in the SM and in the 2HDM are
induced by the up-type quark
self-energy diagrams and are found to be very small.
However, in the SSM, the asymmetry ${\cal A}_b$ can be substantial,
${\cal O}(\alpha_{QCD})$, provided the
CP-violating phase of gluino-top-stop couplings is not suppressed.
Within the SSM ${\cal A}_b$ is generated by the vertex corrections.
\vspace*{2.0cm}
\begin{flushleft}
\parbox{2.in}
{CERN-TH.6941/93\\
July 1993}
\end{flushleft}
\setcounter{page}{0}
\thispagestyle{empty}
\newpage

\section{Introduction}

A very heavy top quark $t$ can offer a few relevant advantages for the
study of CP violation:
\begin{itemize}
\item{If $m_t>130$ GeV, it would decay before it can form a bound
state~\cite{bigi}; therefore the perturbative description is much more
reliable.}
\item{For the same reason, the spin information of the top quark would not be
diluted by hadronization, so that the spin effects can provide
a very useful tool in searching for CP violation.}
\item{Again, because of the large mass of the top quark, its properties are
sensitive to interactions mediated by Higgs bosons~\cite{lee}.}
\item{The Kobayashi-Maskawa~\cite{km} mechanism of CP violation
is strongly suppressed for the top quark since its mixing with other
generations is very weak; therefore it is sensitive to non-conventional sources
 of CP violation.}
\end{itemize}
Discussion of heavy-quark rate asymmetries has been initiated by the
pioneering article by Bander, Silverman and Soni in Ref.~\cite{pioneer},
which was
devoted to the bottom-quark decays. The same mechanism of the
rate-asymmetry generation was then improved~\cite{rescat} and
eventually applied to the top quark. The top-quark decay-rate
asymmetry has been discussed in the literature~\cite{ratelit},
usually for cases of 3-body decays, which make the problem much more
involved. Here, we will concentrate on dominant 2-body decays of the top
quark: $\twddec$.
Other consequences of CP violation due to the top quark spin effects can be
found in
refs.\cite{cpcol}--\cite{susywerner}.

%
\section{SM and 2HDM}
Let us denote the contribution from $d_i$ internal quark to the mixed
$t$--$u_j$ quark self-energy diagram by
\beq
i\{\not{\!p_t}[P_L\Sigma_L(\mis)+P_R\Sigma_R(\mis)]
+P_Lm_j\Sigma^S_L(\mis)+P_Rm_t\Sigma^S_R(\mis)\},
\label{selfent}
\eeq
where $P_{L/R}$ are the chiral projection operators,
$p_t$ is the incoming top-quark momentum, $m_t$ denotes the top-quark mass,
$p_t^2=\mts$,  and $m_j$ is the mass of the outgoing $u_j$ quark.
{\it Hereafter we assume}
that $V_{ij}$-elements of the Kobayashi-Maskawa (KM) matrix~\cite{km}
are factorized out and do not enter into $\Sigma$'s.
The analogous expression can be written for the incoming antitop
of momentum $p_{\bar t}$;
\beq
i\{\not{\!{p}_{\bar t}}[P_L\bar{\Sigma}_L(\mis)
+P_R\bar{\Sigma}_R(\mis)]+P_Lm_t\bar{\Sigma}^S_L(\mis)
+P_Rm_j\bar{\Sigma}^S_R(\mis)\}.
\label{selfentbar}
\eeq
It is easy to see that the following relations hold:
\beq
\Sigma_{L/R}=-\bar{\Sigma}_{L/R} \;\;\hbox{and}\;\;
\Sigma^S_{L/R}=\bar{\Sigma}^S_{R/L}.
\label{relations}
\eeq
For the tree-level amplitude we use the following notation:
\beq
A_1=V^\ast_{tk} \hat{A}_1 \; , \;\;\;\;\;\;\;
\hat{A}_1= - \frac{ig}{\sqrt{2}}
\bar{u}(p_k)\gamma_\mu P_L u(p_t) \epsilon^\mu,
\label{treeamp}
\eeq
where $\epsilon^\mu$ is the $W$-boson polarization vector. The diagram
in Fig.~\ref{fig:selfcut} with the self-energy inserted before the decay
vertex produces a 1-loop amplitude for the top-quark decay:
\beq
A_2=-{\hat A_1} \sum_{ij}V^\ast_{ti}V^{}_{ji}V^\ast_{jk}\frac{\mt^2}
{\mt^2-\mj^2}\Sigma(\mis),
\eeq
where $i$ denotes the internal down-type quark flavour and
\beq
\Sigma(\mis) \equiv
\Sigma_L(\mis)+\Sigma^S_R(\mis)+\frac{\mj}{\mt}\Sigma_R(\mis)
+\frac{\mjs}{\mts}\Sigma^S_L(\mis).
\label{sigma}
\eeq
%
\begin{figure}[ht]
\caption
{The self-energy induced contribution to
${\cal A}_k$. The vertical dashed line
denotes the  absorptive part of the corresponding integral.}
\label{fig:selfcut}
\end{figure}
\bigskip
A similar expression can be obtained for the antitop decay. The
rate difference between top and antitop decay is given by the
interference of the tree-level and 1-loop amplitudes:
\beq
2m_t(\Gamma-\bar{\Gamma})=2\int d\Phi(W^+d_k)
[\re(A_1A^\ast_2)-\re(\bar{A}_1\bar{A}^\ast_2)],
\label{ratediff}
\eeq
where $\Gamma \equiv \Gamma(\twddec)$, $\bar{\Gamma} \equiv \Gamma(\tbwddec)$,
$\bar{A}_1$ and $\bar{A}_2$ stands for the tree and 1-loop induced amplitudes
for  the antitop decay.
For the phase-space element $d\Phi(\cdots)$, we use:
\beq
d\Phi=(2\pi)^4 \prod^{n_f}_{i=1} \frac{d^3p_i}{(2\pi)^3 2E_i}
\delta^{(4)}(P_{in}-\sum^{n_f}_{i=1}p_i),
\label{phase}
\eeq
As is usual for rate asymmetries,
since CP-violating couplings enter with opposite signs for the top and
the antitop decays, the asymmetry can be non-zero if there is an
absorptive part of the 1-loop contribution to the amplitude:
\beq
{\cal A}_k \equiv\frac{\Gamma-\bar{\Gamma}}{\Gamma+\bar{\Gamma}}=
-2\sum_{ij}\frac{\imag}{|V_{tk}|^2}\frac{\mts}{\mts-\mjs}
\im[\Sigma(\mis)],
\label{asres}
\eeq
where the relations of Eq.~(\ref{relations}) have been adopted.
The above formula has a few relevant features. Since in the SM there is
only one CP-violating parameter, our asymmetry must be proportional to
it; in fact  all non-zero values of $\imag$ satisfy:
\beq
\imag=\pm J.
\label{cpviol}
\eeq
The $J$ factor, which is invariant
under reparametrization~\cite{cecylia}, can be rewritten
in terms of the Wolfenstein parameters~\cite{wolf} of the KM matrix:
\beq
J=A^2\lambda^6 \eta,
\label{wolfen}
\eeq
where $A \simeq 1$, $\lambda=\sin\theta_{C}=0.22$, and $\eta$ is
a CP-violating parameter of the KM matrix. Because of the unitarity of the
KM matrix, it is clear that non-zero contributions to the sum in
Eq.~(\ref{asres}) appear only if $i \ne k$ and $j \ne t$. This
supports the common belief~\cite{rescat} that rescattering of the
initial (or final) state does not contribute to rate asymmetries.
The last important remark is that the asymmetry is doubly GIM-suppressed;
first because the loop is summed over $i$ and second,
because it is also summed over $j$. Therefore we expect
\beq
{\cal A}_k \sim \lambda^6 \frac{(m^2_{u_l}-m^2_{u_m})}{\mws}
\frac{(m^2_{d_n}-m^2_{d_o})}{\mws}.
\label{asest}
\eeq
This turns out to be a very small number. The asymmetry within
the SM may only be of an academic interest; however it is instructive to
note what the suppression factors are, in order to overcome them within
some modifications of the KM mechanism of CP violation. It also
illustrates many generic features of CP violation. For completeness we
present here the result for ${\cal A}_k$ calculated within the 2HDM under the
assumption that decays $\thddec$ are kinematically allowed.
The relevant contributions to the amplitudes are given by similar diagrams as
in
Fig.~\ref{fig:selfcut},
with the internal $W$ line replaced by the charged Higgs boson line.
We do not consider the possibility that CP is violated through
interactions of neutral Higgs bosons and we keep only contributions
growing with $\tan^2 \beta \equiv (v_2/v_1)^2$:
\beq
{\cal A}_k=-\frac{\alpha}{8\sin^2\theta_W}[f_W(y)+\tan^2\beta f_H(x,y)]
\sum_{ij}\frac{\imag}{|V_{tk}|^2}
\frac{m^2_i}{m^2_W}\frac{m^2_j}{m^2_W},
\label{ashiggs}
\eeq
where the leading terms in the expansion $m_{i,j}^2/m^2_W$ have been kept.
The $f_W$ and $f_H$ are functions originating from imaginary
parts of self-energies involving $W$ and $H$ exchange, respectively;
they are given by:
\bea
f_W(y)=-3(1+y^4)/y^6 &,& f_H(x,y)=(1-x^2)^2/(x^4y^2)\;,\\
y \equiv \mt/\mw     &,& x \equiv  \mt/ \mh         \;. \nonumber
\label{ffun}
\eea
Eventually, assuming $A=1$, we can write the following expression:
\beq
{\cal A}_k=-\frac{\alpha\eta}{8\sin^2\theta_W}
    \frac{m_b^2m_c^2}{m_W^4}[f_W(y)+\tan^2\beta f_H(x,y)]
\left\{ \begin{array}{ccc}
   -[(1-\rho)^2+\eta^2]^{-1} & \mbox{for} & k=d\\
    \lambda^2                & \mbox{for} & k=s\\
   -\lambda^6 m_s^2/m_b^2    & \mbox{for} & k=b.
\eaa \right.
\eeq
{}From the above we see that ${\cal A}_k$ can be, within the SM, at best of the
order of $10^{-9}$, and even that tiny number can only be reached for the
rarest mode: $t \ra W^+ d$. Since $\tanb$ is limited from above by
the measurement~\cite{tanblim} of the branching ratio for $b \ra c \tau
\nu_\tau$ ($\tanb < 0.54 [\mh/1\gev]$) we can conclude that even within
2HDM we are not able to enhance the rate asymmetry.  However, if we
admit more than three generations of quarks ${\cal A}_k$ may become much
bigger,
since in that case we can overcome both sources of suppression: no
$\lambda^6$ factor and
a new $t'$ quark can have a comparable mass as $m_t$ so that the
suppression due to the $u_j$ propagator is absent.
\section{Supersymmetric Standard Models}
In this section we will restrict ourselves to the dominant top-quark decay
mode:
$\twbdec$ and effectively neglect any flavour mixing.
Supersymmetric versions~\cite{cpsusy}
of the SM offer new sources of CP violation.
We write down the CP-violating interaction that is relevant for us:
\bea
{\cal L} &=& i\surd 2 g_s
[{\tilde t}^\ast_L T^a (\bar {\lambda}^a t_L)+
{\tilde t}^\ast_R T^a (\bar {\lambda}^a t_R)] + (t\leftrightarrow b)
\nonumber\\
&&-\;\;(i/\sqrt{2})V_{tb}
\tilde b_L^\dagger
\stackrel{\leftrightarrow}{\partial}_\mu
\tilde t_L W^{-\mu}
+ \hc,
\label{glulag}
\eea
where $g_s$ is the QCD coupling constant.
For our purpose the most relevant new source of CP violation, which
appears in the SSM, would be the phase in the $\tilde t_L-\tilde t_R$
mixing. The stop quarks of different handedness are related to the
stop quark mass eigenstates ${\tilde t}_+$, ${\tilde t}_-$ through the
following transformations:
\bea
\st_L & = & \cos \alpha_t \st_- - e^{i\phi_t}\sin \alpha_t\st_+
\nonumber \\
\st_R & = & e^{-i\phi_t} \sin \alpha_t \st_- + \cos \alpha_t\st_+.
\label{mix}
\eea
The only 1-loop diagram responsible for the generation of ${\cal A}_b$ is
shown in Fig.~\ref{fig:vercut}. The reader can convince himself that
there is no CP-violating contributions coming from the $t-t$ self-energy
diagram;
it is again an illustration of the fact that rescattering into itself
does not contribute to rate asymmetries.
\begin{figure}[ht]
\caption
{The vertex correction to ${\cal A}_b$ in the SSM.
The vertical dashed line denotes the
absorptive part of the corresponding integral.}
\label{fig:vercut}
\end{figure}
\bigskip
The $\tilde{b}_L-\tilde{b}_R$ mixing,
which may also provide the necessary phase, has the same structure as
the one for the top sector, with the substitutions: $\phi_t \rightarrow
\phi_b$ and $\alpha_t \rightarrow \alpha_b$. However, if we assume
that the scalar $b$-quarks are almost degenerate, their mixing effect
can be neglected. The generalization is obvious.
It is worthwhile to observe that, if we add phases $e^{i\phi_\lambda}$
and $e^{-i\phi_\lambda}$ to the terms $(\bar\lambda^a t_L)$ and
$(\bar\lambda^a t_R)$ in Eq.~(\ref{glulag}), owing to the complex gluino mass,
their effect can be absorbed into $\phi_t(\to \phi_t-2\phi_\lambda)$
and $\phi_b(\to \phi_b-2\phi_\lambda)$.
Since the same interactions generate the neutron's electric dipole moment
(NEDM), we have to take into account the limits originating  from this
measurement. However, direct restrictions on $\phi_{t/b}$ from the NEDM
turn out not to be very reliable~\cite{gavela} and therefore will not be
applied here. Indirect bounds may be obtained within the
supergravity-induced SSM. However, as showed in Ref.~\cite{kizukuri},
even assuming
the same phase for all quark families, the model allows for maximal
CP-violating phases for sufficiently heavy  up- and down-squarks;
therefore it is legitimate to assume maximal CP violation. It should be
stressed here that we are not restricting ourselves to the minimal
supergravity-induced models.

A very convenient way to parametrize the $t \to bW^+$
and $\bar t\to \bar b W^-$ decay vertices is the following:
\begin{eqnarray}
\Gamma^\mu & = & \frac{-igV_{tb}^{KM}}{\surd 2}\bar{u}(p_b)\left[
\gamma^\mu  P_L -\frac{2p_t^\mu}{m_W} F^R_2 P_R\right] u(p_t)
\label{decays}
,\\
\bar{\Gamma}^\mu & = & \frac{-ig{V_{tb}^{KM}}^*}
{\surd 2}\bar{v}(p_{\bar{t}})\left[
\gamma^\mu P_L+\frac{2p_{\bar t}^{\mu}}{m_W}
\bar{F}^L_2 P_L\right]  v(p_{\bar{b}}).
\end{eqnarray}
No other relevant form factor is generated at the 1-loop
level of the perturbation expansion.
We relate the rate asymmetry and the form factors as follows:
\begin{equation}
{\cal A}_b \equiv \frac{\Gamma(\twbdec)-\Gamma(\tbwbdec)}
                     {\Gamma(\twbdec)+\Gamma(\tbwbdec)}
={{(m_t/m_W)w^2(m_t^2,m_b^2,m_W^2) \hbox{Re}(F_2^R-\bar{F}_2^L)} \over
{2m_W^4-m_W^2m_t^2-m_W^2m_b^2-(m_t^2-m_b^2)^2}}  \;.
\label{asymm}
\end{equation}
Here the kinematic function
$w(x,y,z)=(x^2+y^2+z^2-2xy-2xz-2zy)^{1\over2}$.
In the SSM, the CP-violating factor, $\hbox{Re}(F_2^R-\bar{F}_2^L)$, is
\footnote{Form factors defined in our previous paper~\cite{r:kg} are related
to those adopted here by the relations: $f_2^{L/R}=F_2^{L/R}$,
$\bar{f}_2^{L/R}=\bar{F}_2^{L/R}$, {\it etc.}
Therefore one can see that for
CP-violating contributions we get
$(F_2^R-\bar{F}_2^L)=(f_2^R-\bar{f}_2^L)=2{f_2^R}_{CPV}$,
where in the last step Eq.(3) of Ref.~\cite{r:kg} has been used.
It is clear how the CP-conserving pieces cancel.}
\begin{equation}
\hbox{Re} (F_2^R-\bar{F}_2^L)=
-\frac{2\alpha_s}{3\pi}\sin(2\alpha_t) \sin(\phi_t)
\frac{m_\lambda m_W}{(m_t^2-m_W^2)^2}
\hbox{Im } {\cal I}(m^2_{\tilde t}) .
\label{resform}
\end{equation}
%
In this article, we assume that only the lightest stop quark $\tilde t_+$,
of  mass $m_{\tilde t}$, is available in the absorptive integral
$\hbox{Im } {\cal I}(m^2_{\tilde t})$. We have
\begin{equation}
\hbox{Im } {\cal I}(m^2_{\tilde t}) /\pi =
   (1+m_W^2/m_t^2)w(m_t^2,m_\lambda^2,m_{\tilde t}^2)
-DL/w(m_t^2,m_b^2,m_W^2)
\;;
\nonumber
\end{equation}
$$D=(m_W^2-m_t^2)^2-(m_{\tilde t}^2 -m_\lambda^2-m_t^2)(m_W^2-m_t^2)
 -(m_{\tilde b}^2-m_\lambda^2)(m_t^2+m_W^2)$$
$$L=\log(L^+/L^-);  \quad
L^\pm=F \pm w(m_t^2,m_W^2,m_b^2)w(m_t^2,m_{\tilde t}^2,m_\lambda^2) \;;$$
$$F=2m_t^2(m_\lambda^2+m_b^2-m_{\tilde b}^2)
         -(m_t^2-m_{\tilde t}^2+m_\lambda^2)
          (m_t^2+m_b^2-m_W^2) \;. $$
\begin{figure}[hbt]
\caption
{The rate asymmetry ${\cal A}_b$ as a function of
the scalar bottom quark mass $m_{\tilde b}$ for various $m_t$
when $m_\lambda=100$ GeV and $m_{\tilde t}=50$ GeV.}
\label{fig:asym}
\end{figure}
\bigskip
In Fig.~3, we show a typical size of the rate asymmetry ${\cal A}_b$ at
the level of one per cent, for general cases.
%
The best environment for measuring the rate asymmetry would mostly
likely be the future linear $e^+e^-$ collider at $500 \gev$.
Since it is expected that about $50000$ $t\bar{t}$ events will be
produced each year, the smallest possible measurable
asymmetry will be of the order of $1\:\%$, assuming that
$BR(\twbdec) \simeq 50\:\%$, for one year's run.
Therefore we conclude that the measurement of the rate asymmetry
induced within the SSM will be quite challenging.
Noticing that if $\ttbar$ are pair produced, nonvansihing
asymmetry must be an effect of CP violation in the process of
$\ttbar$ decays because whenever $t$ is being produced there is an
accompanied $\bar t$ and therefore the production mechanism cannot fake
the asymmetry in the decay rates.
%
\section{CPT Constraints}
The CPT invariance guarantees that
the particle and the antiparticle have the same total widths. Assuming no
cancellation between different orders of perturbation expansion, we can
therefore conclude that, order by order, contributions to the total
widths should be the same. In particular this means that at the order of
$\alpha^2$:
\beq
\sum_{\mbox{decay channels}}\Gamma(t \ra \cdots)=
\sum_{\mbox{decay channels}}\Gamma(\tbar \ra \overline{\cdots}).
\label{cpt}
\eeq
One may have noticed that besides 2-body decays considered earlier
there are also 3-body
decays contributing at this level
of perturbation expansion. However their effect is identical
channel by channel,
for $t$ and $\bar{t}$ and therefore irrelevant for our purpose.
It is related to the fact that the $W$ self-energy amplitudes
do not induce any rate asymmetry.

It is worthwhile to observe how the above constraint is fulfilled.
Let us restrict ourself to the SM. In
the case of self-energy induced asymmetries, it is particularly simple;
one can see from the formula for the asymmetry, Eq.~(\ref{asres}), that
the rate difference summed over all 2-body final states (summation over
$k$) could be written as:
\begin{eqnarray}
&&\sum_k[\Gamma(\twddec)-\Gamma(\tbwddec)] \nonumber\\
&\sim&
\sum_k \int d\Phi(Wd_k)\frac{1}{2}|\hat{A}_1|^2 \sum_{ij}
\imag \im[\Sigma(\mis)]  \;.
\label{asselfnew}
\end{eqnarray}
An irrelevant factor from the $j$ propagator is not shown here.
Since
\beq
\int d\Phi(W^+d_k)\frac{1}{2}|\hat{A}_1|^2=\mt \im[\Sigma(m^2_k)]    \;,
\label{dispself}
\eeq
we can write the rate difference summed over all 2-body final states as
\beq
\sum_k[\Gamma(\twddec)-\Gamma(\tbwddec)]\sim
\sum_{ijk}\imag \im[\Sigma(\mis)] \im[\Sigma(m^2
_k)]=0 \;.
\label{summedas}
\eeq
It vanishes because $\imag$ is antisymmetric under $i \leftrightarrow k$
exchange. In the 2HDM, if $\thddec$ is kinematically allowed one should
include also this decay channel, the conclusion stays, of course, the
same.

A similar situation occurs for the vertex-induced asymmetry within the SSM.
Since we neglect any flavour mixing,
there are only two 2-body decay channels opened: $\twbdec$ and
$\tsgdec$. It is obvious from Eq.~(\ref{cpt}) that at least two decay
channels must exist in order to generate a non-zero rate asymmetry.
Equation~(\ref{cpt}) tells us also that the following relation is
satisfied:
\beq
[\Gamma(\twbdec) - \Gamma(\tbwbdec)]=- [\Gamma(\tsgdec) - \Gamma(\tbsgdec)]
\;.
\label{aseq}
\eeq
It is convenient to adopt here a phase convention in which CP violation
in the process $\twbdec$ appears only in the $(t  \tilde{t}_+  \lambda)$
vertex.
%
%
In such basis it is easy to separate out
CP-violating phases in the tree-level amplitude
$A_1(t\to\tilde t \lambda)$ and the 1-loop amplitude $A_2(\twbdec)$.
It will be convenient to adopt the following definitions for
CP-violating contributions to those amplitudes:
\beq
A_2(\twbdec)=e^{i\phi}\hat{A}_2(\twbdec)\;,\;\;\;\;
A_1(\tsgdec)=e^{i\phi}\hat{A}_1(\tsgdec).
\eeq
In order to see how the CPT-constraint is satisfied, it is
instructive to write down an expression for the asymmetry before the
loop integration is performed, by means of the Cutkosky~\cite{cut} rule
for the appropriate absorptive part:
\bea
\lefteqn{[\Gamma(\twbdec) - \Gamma(\tbwbdec)] \sim } \\
&&-2\im(e^{i\phi}) \int d\Phi(\st \glu) \int d\Phi(\wpp b)
\hat{A}_1(\tsgdec) A(Wb \ra \st \glu)^\ast A(\twddec)^\ast .
\label{befint}
\eea
For formal reasons we put stars to indicate complex conjugations,
although both $A(Wb \ra \st \glu)$ and $A(\twddec)$ are real numbers;
$\int d\Phi(\cdots)$
represents an appropriate phase-space integration including summation
over spins of on-shell particles. The reader can easily verify
that an analogous rate difference for $\tsgdec$ would have an
opposite sign.
\section{Summary}
We have discussed the CP violation in the decay  rate-asymmetry of the
top quark in various models. We found  effect at a
level of a per cent in the SSM, where a CP-violating phase may occur in the
$\tilde t_L$--$\tilde  t_R$ mixing. We also illustrate how the CPT
constraint manifests itself  in explicit calculations such that
the total widths of a top quark and an antitop quark are equal.

\vspace{1cm}
\centerline{\bf Acknowledgments}
\vspace{.5cm}

We thank J. Liu for a useful discussion.
We are grateful to the CERN Theory Division for the support during this work.
This research was also supported in part by the U.S. Department of Energy.

\end{document}